\documentclass[nobibnotes,nofootinbib,aps,pre,showpacs,
preprint
]{revtex4}

\usepackage{setspace}
\usepackage{graphics}
\usepackage{graphicx}
\usepackage{color}
\usepackage{amsmath}
\usepackage{amssymb}
\usepackage{bm}
\usepackage{float}
\usepackage{epsfig}
\usepackage{epstopdf}
\usepackage{color,graphics}
\usepackage[toc,page]{appendix}

\begin{document}
\title{Neutrino-driven electrostatic instabilities in a magnetized plasma}

\author{Fernando Haas and Kellen Alves Pascoal}
\affiliation{Instituto de F\'{\i}sica, Universidade Federal do Rio Grande do Sul, Av. Bento Gon\c{c}alves 9500, 91501-970 Porto Alegre, RS, Brasil}

\author{Jos\'e Tito Mendon\c{c}a}
\affiliation{IPFN, Instituto Superior T\'ecnico, Universidade de Lisboa, 1049-001 Lisboa, Portugal}
\affiliation{Instituto de F\'isica, Universidade de S\~ao Paulo, 05508-090 S\~ao Paulo, SP, Brasil}

\begin{abstract}
The destabilizing role of neutrino beams on the Trivelpiece-Gould modes is considered, assuming electrostatic perturbations in a magnetized plasma composed by electrons in a neutralizing ionic background, coupled to a neutrino species by means of an effective neutrino force arising from the electro-weak interaction. The magnetic field is found to significantly improve the linear instability growth rate, as calculated for Supernova type II environments. On the formal level, for wave vector parallel or perpendicular to the magnetic field the instability growth rate is found from the unmagnetized case replacing the plasma frequency by the appropriated Trivelpiece-Gould frequency. The growth rate associated with oblique propagation is also obtained. 
\end{abstract}

\maketitle

\section{Introduction}

There is a continuous interest on the neutrino-plasma interaction in magnetized media. For instance, it has been suggested 
\cite{Bethe1}--\cite{Bludman} that neutrino bursts could transfer energy-momentum to the magnetized plasma around the core of the supernovae, triggering the stalled shock expansion therein. Strong wakefields driven by neutrino bursts in magnetized electron-positron plasma have been reported \cite{mag2}. The Mikheilev-Smirnov-Wolfenstein effect of neutrino flavor conversion is significantly influenced by strong magnetic fields, with possible implications on supernova evolution and other magnetized media \cite{mag3}. Spin waves destabilized by neutrino beams in magnetized plasma \cite{Semikoz}, the linear spectrum in magnetized electron–positron coupled to neutrino-antineutrino species in the early universe and neutrino cosmology \cite{mag5}, the neutrino effective charge in magnetized pair plasma \cite{mag1}, neutrino emission via collective processes in magnetized plasma \cite{mag7}, nonlinear generation of waves by neutrinos in magnetized plasmas \cite{mag8, mag9}, the neutrino destabilizing effects on magnetosonic waves described by neutrino magnetohydrodynamics model \cite{nmhd} and the coupling between neutrino flavor oscillations and ion-acoustic waves \cite{pre} have been reported. 
In astrophysical plasmas in general, intense neutrino beams are ubiquitous, as in the lepton era of the early universe \cite{Tajima}.

Trivelpiece-Gould modes \cite{Trivelpiece} are one of the basic waves in magnetized plasma, characterized by electrostatic excitations only (no magnetic field perturbations), for an electron plasma in an homogeneous ionic background. Therefore, the treatment of Trivelpiece-Gould modes allowing for neutrino-plasma coupling has an intrinsic relevance, besides astrophysical applications. The solution of the problem was not performed yet and this is the goal of the work. Notice that according to the original article \cite{Trivelpiece}, Trivelpiece-Gould modes were deduced allowing for arbitrary angle between the external magnetic field and wave vector, see also e.g. \cite{Chen} (p. 107). 

The article is organized as follows. In Section II the basic model equations are proposed. In Section III the general dispersion relation is obtained. Section IV treats two notable subcases: wave propagation perpendicular and parallel to the external magnetic field. The destabilization and growth rate of the corresponding Trivelpiece-Gould modes is then derived and calculated in astrophysical scenarios. Section V contains the oblique propagation case. Section VI has our conclusions. Appendix A is reserved to the complete expressions of the neutrino number density and velocity field perturbations.

\section{Physical Model}

The system is described by an hydrodynamical model for electrons and neutrinos, in an homogeneous ionic background. Denoting $n_{e,\nu}$ and ${\bf u}_{e,\nu}$ as respectively the electron ($e$) and neutrino ($\nu$) fluid densities (in the laboratory frame) and velocity fields, one will have the continuity equations 
\begin{equation}
\frac{\partial n_e}{\partial t} + \nabla \cdot (n_e {\bf u}_e) = 0 \,,  \quad \frac{\partial n_\nu}{\partial t} + \nabla \cdot (n_\nu {\bf u}_\nu) = 0 \,, \label{eq01} 
\end{equation}
together with the (non-relativistic) electron force equation
\begin{equation}
m_e\left(\frac{\partial}{\partial t} + {\bf u}_e\cdot \nabla \right)\,{\bf u}_e = - \,\frac{\nabla p}{n_e} - e\left(- \,\nabla\phi + {\bf u}_e \times {\bf B}_0\right) + \sqrt{2}\,G_F\,({\bf  E}_\nu + {\bf  u}_{e}\times{\bf  B}_\nu) \,, \label{eq02}
\end{equation}
and the neutrino force equation 
\begin{eqnarray}
 \frac{\partial {\bf p}_\nu}{\partial t} + {\bf u}_\nu \cdot \nabla {\bf p}_\nu &=& \sqrt{2}\,G_F \left(
- \nabla n_e - \frac{1}{c^2}\,\frac{\partial}{\partial t}\,(n_e{\bf u}_e) + \frac{{\bf u}_\nu}{c^2} \times \left[\nabla\times(n_e{\bf u}_e)\right]
\right) \,, \label{eq07} 
\end{eqnarray}
where ${\bf p}_\nu = \mathcal{E}_\nu {\bf u}_\nu/c^2$ is the neutrino relativistic momentum for a neutrino beam energy $\mathcal{E}_\nu$. In Eq. (\ref{eq02}), $m_{e}$ is the electron mass, $- e$ is the electron charge, $p = p(n_e)$ is the electron fluid pressure, $G_F$ is Fermi's coupling constant, and ${\bf  E}_\nu, {\bf  B}_\nu$ are effective neutrino electric and magnetic fields given by
\begin{equation}
 {\bf E}_\nu = - \nabla n_\nu - \frac{1}{c^2}\,\frac{\partial}{\partial t}\,(n_\nu {\bf  u}_\nu) \,, \quad {\bf B}_\nu = \frac{1}{c^2}\,\nabla  \times (n_\nu {\bf  u}_\nu) \,, \label{eq04}
\end{equation}
where $c$ is the speed of light. In this work we consider electrostatic excitations with scalar potential $\phi$ described by Poisson's equation with a neutralizing background $n_0$, 
\begin{equation}
\label{poi}
\nabla^2\phi = \frac{e}{\varepsilon_0}\,(n_e - n_0) \,,
\end{equation}
where $\varepsilon_0$ the vacuum permittivity constant, in the presence of an homogeneous magnetic field ${\bf B}_0$ as apparent in the magnetic force in Eq. (\ref{eq02}). However, there are no magnetic field perturbations. Without neutrinos, this setting gives rise to the Trivelpiece-Gould modes \cite{Trivelpiece}. Our goal is to investigate the role of a neutrino beam free energy in this context. The present model was introduced, without ambient magnetic field, in \cite{Serbeto}. For simplicity, neutrino flavor oscillations are not taken into account.

\section{Linear waves}

We have the homogeneous static equilibrium 
\begin{eqnarray}
n_e &=&  n_0 \,, \quad {\bf u}_e = 0 \,, \quad n_\nu = n_{\nu 0} \,, \quad {\bf u}_\nu = {\bf u}_{\nu 0} \,, \quad \phi = 0 \,,
\end{eqnarray}
where $n_{\nu 0}$ and ${\bf u}_{\nu 0}$ are respectively the equilibrium neutrino number density and velocity field, assumed to be constant. Linearizing the model equations in terms of plane wave perturbations $\sim \exp[i({\bf k}\cdot{\bf r} - \omega t)]$, denoting fluctuations with a delta as for instance in $n_e = n_0 + \delta n_e \exp[i({\bf k}\cdot{\bf r} - \omega t)]$, one readily find 
\begin{eqnarray}
\omega\,\delta n_e &=& n_0 \,{\bf k}\cdot\delta{\bf u}_e \,, \quad (\omega - {\bf k}\cdot{\bf u}_{\nu 0})\,\delta n_\nu = n_{\nu 0} \,{\bf k}\cdot\delta{\bf u}_\nu \,, \label{A} \\
m_e \,\omega \,\delta{\bf u}_e &=& \frac{1}{n_0}\left(\frac{dp}{dn_e}\right)_0\,{\bf k}\,\delta n_e - e\, ({\bf k}\,\delta\phi + i \,\delta{\bf u}_e\times{\bf B}_0) \nonumber \\ &+& \sqrt{2}\, G_F \left(({\bf k} - \frac{\omega}{c^2}{\bf u}_{\nu 0})\,\delta n_\nu - \frac{\omega n_{\nu 0}}{c^2} \delta{\bf u}_\nu\right) \,, \label{B} \\
(\omega - {\bf k}\cdot{\bf u}_{\nu 0})\delta{\bf p}_\nu &=& \sqrt{2} G_F \left({\bf k} \delta n_e - \frac{n_0 \,\omega}{c^2}\,\delta{\bf u}_e - \frac{n_0}{c^2}\,{\bf u}_{\nu 0}\times({\bf k}\times\delta{\bf u}_e)\right) \,, \label{C} \\
-k^2 \delta\phi &=& \frac{e}{\varepsilon_0}\,\delta n_e \,. \label{D}
\end{eqnarray}

Notice that in Eq. (\ref{C}), $\delta{\bf u}_e$ appears already in a term proportional to $G_F$. Since there is no need to include very small higher order corrections, in Eq. (\ref{C}) we need only the classical $\delta{\bf u}_e = \delta{\bf u}_{e}^C$ obtained setting $G_F = 0$ in Eq. (\ref{B}), namely, 
\begin{equation}
\delta{\bf u}_e^{C} = \frac{\delta n_e}{n_0}\frac{V^2}{\omega (\omega^2 - \omega_c^2)} \left(\omega^2\, {\bf k} - ({\bf k}\cdot\boldsymbol{\omega}_c)\,\boldsymbol{\omega}_c
+ i \,\omega \,\boldsymbol{\omega}_c\times{\bf k}\right) \,, \label{trick}
\end{equation}
where 
\begin{equation}
V^2 = v_{T}^2 + \frac{\omega_{p}^2}{k^2} \,, \quad v_{T}^2 = \frac{1}{m_e}\left(\frac{dp}{dn_e}\right)_0 \,, \quad \omega_p^2 = \frac{n_0 e^2}{m_e \varepsilon_0} \,, \quad \boldsymbol{\omega}_c = \frac{e{\bf B}_0}{m_e} \,.
\end{equation}
The trick is to substitute $\delta{\bf u}_e \rightarrow \delta{\bf u}_e^C$ in Eq. (\ref{C}), to obtain $\delta{\bf p}_\nu$ and then $\delta{\bf u}_\nu$ correct up to ${\cal O}(G_F)$. Using the neutrino continuity equation, this will give $\delta n_\nu$ also up to ${\cal O}(G_F)$. The recursive procedure allows to rewrite Eq. (\ref{B}) as 
\begin{equation}
\label{x}
\omega\, \delta{\bf u}_e + i \,\delta{\bf u}_e \times \boldsymbol{\omega}_c = \frac{V^2 \,{\bf k}\, \delta n_e}{n_0} + \omega \,\delta{\bf v}_\nu \,,
\end{equation}
where $\delta{\bf v}_\nu$ contains all neutrino effects, 
\begin{equation}
\delta{\bf v}_\nu \equiv \frac{\sqrt{2}\, G_F}{m_e \omega}\left(({\bf k} - \frac{\omega}{c^2}{\bf u}_{\nu 0})\,\delta n_\nu - \frac{n_{\nu 0}\, \omega}{c^2}\,\delta{\bf u}_\nu\right) \,.
\end{equation}
By construction, $\delta{\bf v}_\nu$ will be of order ${\cal O}(G_{F}^2)$, since $\delta n_\nu$ and $\delta{\bf u}_\nu$ are ${\cal O}(G_{F})$ by the procedure, whose ultimate expressions are shown in Appendix A. The same formulae show $\delta n_\nu$ and $\delta{\bf u}_\nu$ as directly proportional to $\delta n_e$. The solution to Eq. (\ref{x}), 
\begin{equation}
\label{O2}
\delta{\bf u}_e = \delta{\bf u}_e^C + \frac{1}{(\omega^2 - \omega_c^2)} \left(\omega^2 \delta {\bf v}_\nu - (\boldsymbol{\omega}_c\cdot\delta{\bf v}_\nu)\,\boldsymbol{\omega}_c + i \,\omega \,\boldsymbol{\omega}_c\times\delta{\bf v}_\nu\right) 
\end{equation}
yields $\delta{\bf u}_e$ proportional to $\delta n_e$ and valid up to ${\cal O}(G_F^2)$. Finally, substituting Eq. (\ref{O2}) into the electrons continuity equation, one derive the linear dispersion relation of Trivelpiece-Gould modes modified by a neutrino beam. As a remark, note that in Eq. (\ref{trick}) and afterward it is assumed $\omega^2 \neq \omega_c^2$, with no real loss of generality since the possible mode with $\omega^2 = \omega_{c}^2$ is neutrino-independent, see Section IVb. 

Proceeding as explained 
gives
\begin{eqnarray}
\delta{\bf p}_\nu &=& \frac{\sqrt{2} G_F \delta n_e}{(\omega - {\bf k}\cdot{\bf u}_{\nu 0})(\omega^2 - \omega_c^2)} \times \nonumber \\ &\times& \Bigl[(\omega^2 - \omega_c^2){\bf k} 
- \frac{V^2}{c^2}\Bigl(\omega^2 {\bf k} - ({\bf k}\cdot\boldsymbol{\omega}_c)\, \boldsymbol{\omega}_c + i (\omega - {\bf k}\cdot{\bf u}_{\nu 0})\boldsymbol{\omega}_c\times{\bf k} \label{x1} \\ &-& \frac{{\bf k}\cdot\boldsymbol{\omega}_c}{\omega}\,{\bf u}_{\nu 0}\times({\bf k}\times\boldsymbol{\omega}_c) + i {\bf k}\,[{\bf u}_{\nu 0}\cdot(\boldsymbol{\omega}_c\times{\bf k})]\Bigr)\Bigr] \,.  \nonumber
\end{eqnarray}

On the other hand, the neutrino velocity perturbation is derived from $\delta{\bf p}_\nu$ according to  
\begin{equation}
\label{x2}
\delta{\bf u}_\nu = \frac{c^2}{{\cal E}_{\nu 0}}\left(\delta{\bf p}_\nu - \frac{{\bf u}_{\nu 0}\cdot\delta{\bf p}_\nu}{c^2}{\bf u}_{\nu 0}\right) \,,
\end{equation}
as found from the relativistic energy-momentum relation, where ${\cal E}_{\nu 0}$ is the zero-order neutrino beam energy. Using Eqs. (\ref{x1}) and (\ref{x2}) we derive a long expression for $\delta{\bf u}_\nu$, which in turn  gives $\delta n_\nu$ from Eq. (\ref{A}). These expressions are shown in the Appendix A, allowing to determine $\delta{\bf v}_\nu$ as proportional to 
$\delta n_e$. 

\begin{figure}[!hbt]
\begin{center}
\includegraphics[width=8.0cm,height=6.0cm]{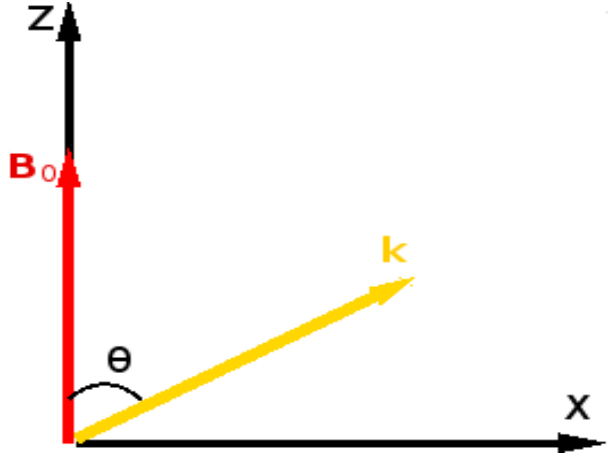}
\end{center}
\caption{Geometry of Trivelpiece-Gould modes.}
\label{figure1}
\end{figure}

Without loss of generality, assuming the ambient magnetic field along the $z-$axis and a wave vector in the $x-z$ plane, as shown in Figure \ref{figure1}, so that 
\begin{equation}
\boldsymbol{\omega}_c = \omega_c \,\hat{z}, \, \quad {\bf k} =  k \,(\sin\theta, 0, \cos\theta) \,. \label{con}
\end{equation}
One then has implicitly the dispersion relation 
\begin{eqnarray}
\left(\omega^4 - \omega_H^2 \omega^2 + \omega_p^2 \omega_c^2 \cos^{2}\theta\right) \delta n_e &=& n_0 \omega \left(\omega^2 {\bf k}\cdot\delta{\bf v}_\nu - ({\bf k}\cdot\boldsymbol{\omega}_c) \,(\boldsymbol{\omega}_c\cdot\delta{\bf v}_\nu) + i \omega {\bf k}\cdot(\boldsymbol{\omega}_c \times \delta{\bf v}_\nu)\right) \nonumber \\
&=& \frac{\sqrt{2} G_F n_0}{m_e c^2} \Bigl[\omega^2 \Bigl(c^2 k^2 - \omega^2\Bigr)\delta n_\nu - c^2 ({\bf k}\cdot\boldsymbol{\omega}_c)^2 \delta n_\nu \nonumber \\ &+& \omega({\bf k}\cdot\boldsymbol{\omega}_c)\,(\boldsymbol{\omega}_c\cdot{\bf u}_{\nu 0})\delta n_\nu + n_{\nu 0}\omega\,({\bf k}\cdot\boldsymbol{\omega}_c)\,(\boldsymbol{\omega}_c\cdot\delta{\bf u}_\nu) \nonumber \\
&-& i \omega^2\,{\bf k}\cdot\Bigl(n_{\nu 0}\,\boldsymbol{\omega}_c\times\delta{\bf u}_\nu + \boldsymbol{\omega}_c\times{\bf u}_{\nu 0}\,\delta n_\nu\Bigr) 
\Bigr] \,, \label{ld}
\end{eqnarray}
in terms of the upper hybrid frequency $\omega_H = \sqrt{\omega_p^2 + \omega_c^2}$. The neutrino continuity equation was used to eliminate 
${\bf k}\cdot{\bf u}_{\nu 0}$. The quantities $\delta{\bf u}_\nu$ and $\delta n_\nu$ are both long expressions proportional to $\delta n_e$ as shown in Eqs. (\ref{a1}) and (\ref{a2}) in the Appendix. Therefore, for $\delta n_e \neq 0$, one obtains the dispersion relation from Eq. (\ref{ld}). 

Without neutrinos ($\delta{\bf v}_\nu \equiv 0$) one would regain the Trivelpiece-Gould dispersion relation 
\cite{Trivelpiece, Chen}, namely $\omega^4 - \omega_H^2 \omega^2 + \omega_p^2 \omega_c^2 \cos^{2}\theta = 0$. For simplicity, at this point it was assumed $\omega_p \gg k v_T$ so that $V \approx \omega_p/k$, yielding a nicer expression for the classical contribution i.e. the left-hand side of Eq. (\ref{ld}). Thermal effects can be recovered through the systematic replacement $\omega_p^2 \rightarrow \omega_p^2 + k^2 v_T^2$. 

We note that in the unmagnetized case ($\omega_c = 0$) using Eq. (\ref{ld}) together with the appropriate special case from Eq. (\ref{a2}) gives the same found in  \cite{Serbeto, Silva1, Silva2}, namely 
\begin{equation}
\omega^2 = \omega_p^2 + \frac{\Delta\,
(c^2 k^2 - \omega_p^2)^2}{(\omega - {\bf k}\cdot{\bf u}_{\nu 0})^2}\times\left(1 - \frac{({\bf k}\cdot{\bf u}_{\nu 0})^2}{c^2 k^2}\right) \,, \label{serb}
\end{equation}
introducing the dimensionless quantity  
\begin{equation}
\Delta = \frac{2\,G_F^2\,n_{0}\,n_{\nu 0}}{m_e\,c^2\,{\cal E}_{\nu 0}} \,. 
\end{equation}
To obtain Eq. (\ref{serb}), in the numerator of the term proportional to $\Delta$ it was replaced the unperturbed approximation $\omega \approx \omega_p$ whenever convenient, since this neutrino term is already a correction. To proceed to the magnetized case, observe that the neutrino contribution in Eq. (\ref{ld}) can be relevant only within a resonance condition where ${\rm Re}(\omega) \approx {\bf k}\cdot{\bf u}_{\nu 0}$, due to the small value of the Fermi constant $G_F = 1.45 \times 10^{-62} \, {\rm J.\,m^3}$. By construction, our calculations retain terms up to ${\cal O}(\Delta)$.

Before embarking in the general case, two subcases are illustrative: wave propagation perpendicular or parallel to the ambient magnetic field, as discussed in the next Section. 

\section{Particular subcases}

\subsection{Wave propagation perpendicular to the ambient magnetic field} 
\label{perpe}

Supposing upper hybrid oscillations with ${\bf k} \perp \boldsymbol{\omega}_c$ and $\omega \neq 0$, one finds from Eqs. (\ref{ld}), (\ref{a1}) and (\ref{a2}),  
\begin{eqnarray}
\omega^2 - \omega_{H}^2 - \Delta \, \omega_{c}^2 &=& \frac{\Delta}{(\omega - {\bf k}\cdot{\bf u}_{\nu 0})^2}\,\Bigl(1 - \frac{\omega^2}{c^2 k^2}\Bigr)\times \nonumber \\
&\times& \Bigl[\Bigl(c^2 k^2 - \omega^2\Bigr)\Bigl(c^2 k^2 - ({\bf k}\cdot{\bf u}_{\nu 0})^2\Bigr) + \Bigl({\bf u}_{\nu 0}\cdot(\boldsymbol{\omega}_c \times {\bf k})\Bigr)^2 \Bigr]  \nonumber \\
&-& \frac{\Delta\, \omega_c^2 (\omega^2 - \omega_H^2)}{\omega^2 - \omega_c^2} + \frac{\Delta\,(c^2 k^2 - ({\bf k}\cdot{\bf u}_{\nu 0})^2)\, \omega^2}{(\omega - {\bf k}\cdot{\bf u}_{\nu 0})^2 (\omega^2 - \omega_c^2)}\Bigl(1 - \frac{\omega^2}{c^2 k^2}\Bigr) (\omega^2 - \omega_H^2) \nonumber \\
&-& \frac{\Delta\, \Bigl({\bf u}_{\nu 0}\cdot(\boldsymbol{\omega}_c \times {\bf k})\Bigr)^2}{(\omega - {\bf k}\cdot{\bf u}_{\nu 0})^2 (\omega^2 - \omega_c^2)}\Bigl(1 - \frac{\omega^2}{c^2 k^2}\Bigr)(\omega^2 - \omega_H^2) \nonumber \\
&+& i\,\frac{\Delta\,\Bigl({\bf u}_{\nu 0}\cdot(\boldsymbol{\omega}_c \times {\bf k})\Bigr)\,\Bigl(c^2 k^2 - ({\bf k}\cdot{\bf u}_{\nu 0})\omega\Bigr)}{(\omega - {\bf k}\cdot{\bf u}_{\nu 0})^2 (\omega^2 - \omega_c^2)}\,(\omega^2 - \omega_H^2) \,.\label{kaka}
\end{eqnarray}
The right-hand side of Eq. (\ref{kaka}) is always a perturbation due to the very small value of the Fermi constant and it is legitimate to replace in it $\omega^2 \rightarrow \omega_{H}^2$ whenever possible and useful. In particular, this substitution allows to discard the explicit imaginary contribution which is proportional to $\Delta\,(\omega^2 - \omega_H^2) \approx 0$ within the accuracy of the approximation. 
The replacement is supported by the numerical results too. We are left with
\begin{eqnarray}
\omega^2 - \omega_{H}^2 - \Delta \, \omega_{c}^2 &=& \frac{\Delta}{(\omega - {\bf k}\cdot{\bf u}_{\nu 0})^2}\,\Bigl(1 - \frac{\omega^2}{c^2 k^2}\Bigr)\times \nonumber \\
&\times& \Bigl[\Bigl(c^2 k^2 - \omega^2\Bigr)\Bigl(c^2 k^2 - ({\bf k}\cdot{\bf u}_{\nu 0})^2\Bigr) + \Bigl({\bf u}_{\nu 0}\cdot(\boldsymbol{\omega}_c \times {\bf k})\Bigr)^2 \Bigr] \,. \label{kperp}
\end{eqnarray}

The non-resonant term $\Delta \, \omega_{c}^2$ on the left-hand side of Eq. (\ref{kperp}) is always very small for realistic conditions, so that it can be dropped too. The right-hand side of the same equation can yield a significant contribution, provided the neutrino beam becomes resonant with the upper-hybrid frequency, so that we set 
\begin{equation}
\omega_H = {\bf k}\cdot{\bf u}_{\nu 0} \,, \quad \omega = \omega_H + \delta \,, \quad |\delta| \ll \omega_H \,, \label{del}
\end{equation}
converting Eq. (\ref{kperp}) into 
\begin{equation}
\omega^2 = \omega_{H}^2 + \frac{\Delta}{(\omega - {\bf k}\cdot{\bf u}_{\nu 0})^2} \times \left[\left(c^2 k^2 - \omega_{H}^2\right)^2 + \left({\bf u}_{\nu 0}\cdot(\boldsymbol{\omega}_c \times {\bf k})\right)^2\right] \times \left(1 - \frac{\omega_H^2}{c^2 k^2}\right) \,, \label{kper}
\end{equation}
which is almost Eq. (\ref{serb}) with the replacement $\omega_p \rightarrow \omega_H$ appropriated to the magnetized case.  

To enhance the neutrino contribution in Eq. (\ref{kper}), ideally one would have $\omega_H \ll c k$. In the non magnetized case, to avoid Landau damping, one also need  $\omega \gg \langle {\bf k}\cdot{\bf v}_e\rangle$, where $\langle\rangle$ denotes the statistical average of the electrons velocities ${\bf v}_e$. 
For almost isotropic electrons equilibrium, it amounts to $\omega \gg k v_T$. This sets \cite{Serbeto, Silva1, Silva2} an upper limit in the wave-number or $k = \omega/v_T$ at which the instability saturates due to electron Landau damping. Although not mandatory, we define $k = \omega/v_T \approx \omega_H/v_T$ in the magnetized case, to access an easier comparison with the unmagnetized results. Notice that now cyclotron  Landau damping is significant for $\omega \approx l \omega_c$, where $l$ is an integer. Such exceptional, damped modes would be described within a kinetic treatment, which is outside the present model.   

In the present context it can be defined 
\begin{equation}
{\bf k} = (k,0,0) \,, \quad \boldsymbol{\omega}_c = (0,0,\omega_c) \,, \quad {\bf u}_{\nu 0} = u_{\nu 0}(\cos\varphi\sin\Theta,\sin\varphi\sin\Theta,\cos\Theta) \,, 
\end{equation}
where for ultra-relativistic neutrinos $u_{\nu 0} \approx c$.  As argued above, setting the wave-number $k \equiv \omega_H/v_T$ transforms Eq. (\ref{kper}) into 
\begin{equation}
\omega^2 = \omega_H^2 + \frac{\Delta \, \omega_H^4 \, c^4/v_T^4}{(\omega - {\bf k}\cdot{\bf u}_{\nu 0})^2} \left[\left(1 - \frac{v_T^2}{c^2}\right)^2 + \frac{\omega_{c}^2 v_T^2}{\omega_H^2 c^2}\,\sin^{2}\varphi\sin^{2}\Theta \right] \,. \label{drr}
\end{equation}
In view of $\omega_c^2 < \omega_H^2$ and the non-relativistic assumption $v_T^2 \ll c^2\,,$ Eq. (\ref{drr}) can be approximated by
\begin{equation}
\omega^2 = \omega_H^2 + \frac{\Delta\, \omega_H^4 \, c^4/v_T^4}{(\omega - {\bf k}\cdot{\bf u}_{\nu 0})^2} \,,
\end{equation}
exactly the same as the non-magnetized result in Eq. (\ref{serb}) for the maximal neutrino perturbation, provided replacing $\omega_p \rightarrow \omega_H$. Moreover, using Eq. (\ref{del}) it is found 
\begin{equation}
\left(\frac{\delta}{\omega_H}\right)^3 = \frac{\Delta}{2} \left(\frac{c}{v_T}\right)^4 \,,
\end{equation}
which corresponds to an unstable mode with 
\begin{equation}
{\rm Im}\left(\frac{\delta}{\omega_H}\right) = \sqrt{3}\,\Delta^{1/3} \left(\frac{c}{2 v_T}\right)^{4/3} > 0 \,. \label{insta}
\end{equation}
Presently the result (\ref{insta}) is the same as the maximal instability growth rate of Refs. \cite{Serbeto, Silva1, Silva2}, with the simple replacement of the plasma frequency by the upper hybrid frequency. Since $\omega_H > \omega_p$, one has an even stronger instability in the magnetized case. Moreover, denoting $\phi$ as the angle between ${\bf k}$ and ${\bf u}_{\nu 0}$, from the resonance condition we find $\cos\phi \approx \omega_H/(c k) \approx v_T/c \ll 1$, showing that the neutrino beam propagates almost perpendicularly to the wave - but without a definite orientation regarding the external magnetic field.

\begin{figure}[!hbt]
\begin{center}
\includegraphics[width=8.0cm,height=6.0cm]{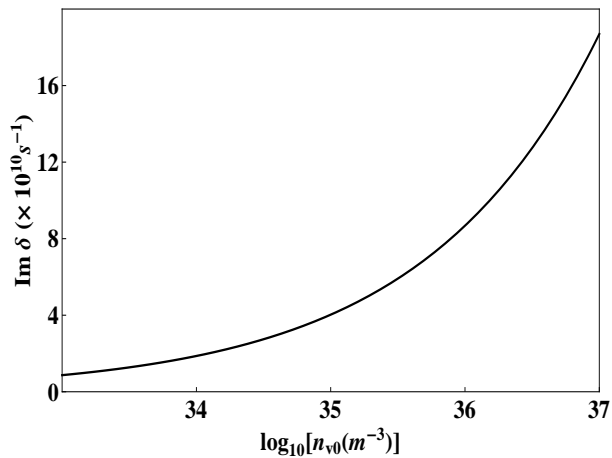}
\end{center}
\caption{Instability growth rate from Eq. (\ref{insta}) for  ${\cal E}_{\nu 0} = 10 \,{\rm MeV}, v_T/c = 1/10, n_0 = 10^{34}\,{\rm m}^{-3}, 
B_0 = 5 \times 10^7 \,{\rm T}$, as a function of neutrino beam number density $n_{\nu 0}$, for ${\bf k} \perp \boldsymbol{\omega}_c$.}
\label{figure2}
\end{figure}

For typical Type II core-collapse scenarios such as for the supernova SN1987A, one has a neutrino burst of $10^{58}$ neutrinos with energies around $10-15$ MeV \cite{Hirata}. To get some estimates, take ${\cal E}_{\nu 0} = 10 \,{\rm MeV}, v_T/c = 1/10, n_0 = 10^{34}\,{\rm m}^{-3}$, appropriate for the center of the star. Moreover, in core-collapse events one has strong magnetic fields $B_0 \approx 10^6 - 10^8 \,{\rm T}$, and we take $B_0 = 5 \times 10^7 \,{\rm T}$. For these parameters, we have $\omega_p = 5.64 \times 10^{18} \,{\rm rad/s}$, a gyrofrequency $\omega_c = 8.78 \times 10^{18} \,{\rm rad/s}$, and $\omega_H = 1.04 \times 10^{19}\, {\rm rad/s}$, showing the salient role of magnetization. The instability growth rate from Eq. (\ref{insta}) is shown in Figure \ref{figure2} as a function of the neutrino beam density $n_{\nu 0}$ 
between $10^{34} - 10^{37} \, {\rm m}^{-3}$. Typically, one has $1/{\rm Im}(\delta) \approx 10^{-11} \,{\rm s}$, to be compared with the characteristic time of supernova explosions, around 1 second. 

\subsection{Wave propagation parallel to the ambient magnetic field} 

When ${\bf k} \parallel \boldsymbol{\omega}_c$, or $\theta = 0^{\circ}$, Eq. (\ref{ld}) simplifies to 
\begin{equation}
(\omega^2 - \omega_c^2) (\omega^2 - \omega_p^2) \delta n_e = \frac{\sqrt{2} G_F n_0}{m_e c^2} (\omega^2 - \omega_c^2) (c^2 k^2 - \omega^2) \delta n_\nu \,. \label{kpara}
\end{equation}
By inspection, the classical mode with $\omega^2 = \omega_c^2$ has no neutrino contribution so that it will be ignored. Therefore we can replace $\omega^2 \approx \omega_p^2 \neq \omega_c^2$ on the right-hand side of Eq. (\ref{kpara}) to obtain 
\begin{equation}
(\omega^2 - \omega_p^2) \delta n_e = \frac{\sqrt{2} G_F n_0}{m_e c^2} (c^2 k^2 - \omega_p^2) \delta n_\nu \,, \label{ser}
\end{equation}
a result which could be directly confirmed from Eqs. (\ref{A}), (\ref{B}) and (\ref{D}). 
Now using Eq. (\ref{a2}) for $\delta n_\nu$, from Eq. (\ref{ser}) we rederive Eq. (\ref{serb}). Therefore for parallel propagation the ambient magnetic field does not modify the instability at all. Proceeding as usual, setting 
\begin{equation}
\omega_p = {\bf k}\cdot{\bf u}_{\nu 0} \,, \quad \omega = \omega_p + \delta \,, \quad |\delta| \ll \omega_p \,,
\end{equation}
the unstable mode is found with 
\begin{equation}
\left(\frac{\delta}{\omega_p}\right)^3 = \frac{\Delta}{2}\,\frac{(1 - \cos^{2}\phi)^3}{\cos^{4}\phi} \,,
\end{equation}
where $\phi$ is the angle between ${\bf k}$ and ${\bf u}_{\nu\,0}$ so that $\omega_p \approx c k \cos\phi$. For parallel propagation (${\bf k} \parallel {\bf B}_0$) the issue of Landau damping becomes relevant for resonant particles gyrating around the magnetic field with the same angular frequency as the wave electric field, ot $\omega - l \omega_c - k v_z \approx 0$, where $l$ is an integer and $v_z$ is the component of the electrons velocity in the direction of ${\bf B}_0$. For the fundamental mode ($l = 0$) and quasi isotropic particle distribution function one then needs $k << \omega_p/v_T$ and so $\cos\phi >> v_T/c$. Finally, one obtains
\begin{equation}
{\rm Im}\left(\frac{\delta}{\omega_p}\right) = \sqrt{3}\,\Delta^{1/3} \, \left(\frac{c}{2 v_T}\right)^{4/3} > 0 \,, \label{lel}
\end{equation}
which is well documented in the literature \cite{Serbeto, Silva1, Silva2} and where $\cos\phi \approx v_T/c << 1$ was selected. In this sense, Eq. (\ref{lel}) is the upper limit of the instability growth rate, avoiding Landau damping. 

\begin{figure}[!hbt]
\begin{center}
\includegraphics[width=8.0cm,height=6.0cm]{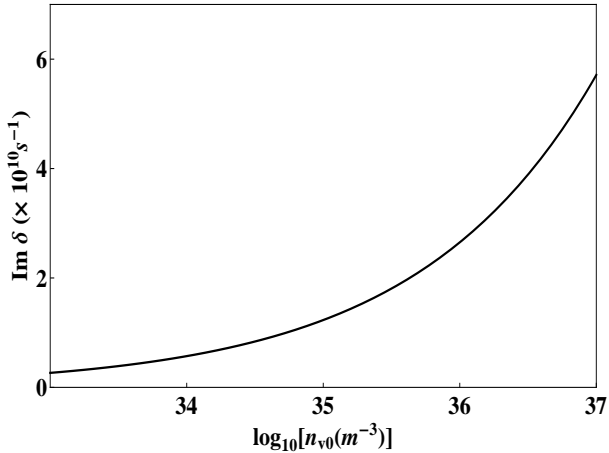}
\end{center}
\caption{Instability growth rate from Eq. (\ref{lel}) for  ${\cal E}_{\nu 0} = 10 \,{\rm MeV}, v_T/c = 1/10, n_0 = 10^{34}\,{\rm m}^{-3}$, as a function of neutrino beam number density $n_{\nu 0}$, for ${\bf k} \parallel \boldsymbol{\omega}_c$.}
\label{figure3}
\end{figure}

It is interesting to compare with the magnetic field dominated case. Using Eq. (\ref{lel}) and exactly the same parameters of subsection \ref{perpe}, one get the result shown in Fig. \ref{figure3}, showing a significantly smaller (but still fast) instability growth rate when compared to Fig. \ref{figure2}. The main conclusion is that a strong ambient magnetic field can have a marked impact on the neutrino-plasma unstable mode, at least for certain wave vector orientations. 

\section{General case} 

For arbitrary angle $\theta$, Eq. (\ref{ld})  becomes more demanding. To start solving it, notice that from inspection of Eqs. (\ref{a1}) and (\ref{a2}) at resonance the terms containing $\delta{\bf u}_\nu \sim (\omega - {\bf k}\cdot{\bf u}_{\nu 0})^{-1}$ in Eq. (\ref{ld}) are generically less singular than those with $\delta n_\nu \sim (\omega - {\bf k}\cdot{\bf u}_{\nu 0})^{-2}$. In this way, dropping the $\delta{\bf u}_\nu$ terms, the linear dispersion relation can be simplified to
\begin{eqnarray}
(\omega^4 - \omega_{H}^2 \omega^2 &+& \omega_p^2 \omega_c^2 \cos^{2}\theta)\,\delta n_e = \frac{\sqrt{2} G_F n_0}{m_e c^2} \Bigl[\omega^2 (c^2 k^2 - \omega^2) - c^2 ({\bf k}\cdot\boldsymbol{\omega}_c)^2 + \nonumber \\ &+& \omega ({\bf k}\cdot\boldsymbol{\omega}_c)(\boldsymbol{\omega}_c\cdot{\bf u}_{\nu 0}) - i \omega^2 {\bf k}\cdot (\boldsymbol{\omega}_c\times{\bf u}_{\nu 0})\Bigr]\delta n_\nu \,. \label{aa2} 
\end{eqnarray}
Moreover, at resonance ($\omega \approx {\bf k}\cdot{\bf u}_{\nu 0}$) it is possible to considerably simplify Eq. (\ref{a2}) as 
\begin{eqnarray}
\delta n_\nu &=& \frac{\sqrt{2} G_F n_{\nu 0} \delta n_e}{{\cal E}_{\nu 0}(\omega - {\bf k}\cdot{\bf u}_{\nu 0})^2(\omega^2 - \omega_{c}^2)} \times \Bigl(1 - \frac{\omega^2}{c^2 k^2}\Bigr) \times \nonumber \\ &\times& \Bigl[(\omega^2 - \omega_c^2) c^2 k^2 - \omega^2 \omega_p^2 + \frac{\omega_p^2}{\omega}({\bf k}\cdot\boldsymbol{\omega}_c)({\bf u}_{\nu 0}\cdot\boldsymbol{\omega}_c)
-i\omega_p^2[{\bf u}_{\nu 0}\cdot(\boldsymbol{\omega}_c\times{\bf k})]\Bigr]  \,. \label{nv}
\end{eqnarray} 

Inserting ({\ref{nv}) into Eq. (\ref{aa2}) and replacing whenever convenient the zero-order expression $({\bf k}\cdot\boldsymbol{\omega}_c)^2 \approx k^2 \omega^2 (\omega_H^2 - \omega^2)/\omega_p^2$ in the neutrino term, it is found after some rearrangements that 
\begin{eqnarray}
\omega^4 - \omega_H^2 \omega^2 &+& \omega_p^2 \omega_c^2 \cos^{2}\theta = \frac{\Delta}{(\omega - {\bf k}\cdot{\bf u}_{\nu 0})^2(\omega^2 - \omega_{c}^2)}\times\Bigl(1 - \frac{\omega^2}{c^2 k^2}\Bigr) \times \nonumber \\ &\times& \Bigl\{\Bigl[\frac{\omega}{\omega_p}\Bigl((\omega^2 - \omega_c^2) c^2 k^2 - \omega_p^2 \omega^2\Bigr) + \omega_p ({\bf k}\cdot\boldsymbol{\omega}_c)({\bf u}_{\nu 0}\cdot\boldsymbol{\omega}_c)\Bigr]^2 + 
\\ &+& \omega_p^2 \omega^2 [{\bf u}_{\nu 0}\cdot(\boldsymbol{\omega}_c\times{\bf k})]^2 \label{ggen} 
+ i c^2 k^2 [{\bf u}_{\nu 0}\cdot(\boldsymbol{\omega}_c\times{\bf k})] \Bigl(\omega^4 - \omega_H^2 \omega^2 + \omega_p^2 \omega_c^2 \cos^{2}\theta\Bigr) \Bigr\} \,. \nonumber 
\end{eqnarray}
As verified, the explicitly imaginary part in Eq. (\ref{ggen}) vanishes in the order of accuracy of the calculation since $\omega^4 - \omega_H^2 \omega^2 + \omega_p^2 \omega_c^2 \cos^{2}\theta = {\cal O}(\Delta)$. Hence the final general dispersion relation reads 
\begin{eqnarray}
\omega^4 &-& \omega_H^2 \omega^2 + \omega_p^2 \omega_c^2 \cos^{2}\theta = \frac{\Delta}{(\omega - {\bf k}\cdot{\bf u}_{\nu 0})^2(\omega^2 - \omega_{c}^2)}\times\Bigl(1 - \frac{\omega^2}{c^2 k^2}\Bigr) \times \label{gen}    \\ &\times& \Bigl\{\Bigl[\frac{\omega}{\omega_p}\Bigl((\omega^2 - \omega_c^2) c^2 k^2 - \omega_p^2 \omega^2\Bigr) + \omega_p ({\bf k}\cdot\boldsymbol{\omega}_c)({\bf u}_{\nu 0}\cdot\boldsymbol{\omega}_c)\Bigr]^2 + \omega_p^2 \omega^2 [{\bf u}_{\nu 0}\cdot(\boldsymbol{\omega}_c\times{\bf k})]^2\Bigr\}  \,. \nonumber 
\end{eqnarray}
Moreover: (a) for ${\bf k} \parallel \boldsymbol{\omega}_c$ it can be used ${\bf u}_{\nu 0}\cdot\boldsymbol{\omega}_c = ({\bf k}\cdot{\bf u}_{\nu 0})\omega_c/k \approx \omega_p \omega_c/k$ in the neutrino term, reducing Eq. (\ref{gen}) to Eq. (\ref{serb}); (b) for ${\bf k} \perp \boldsymbol{\omega}_c$ and with $\omega \approx \omega_H$, Eq. (\ref{gen}) reduces to Eq. (\ref{kper}). 

Despite the fact that the general result encompasses the subcases of Section IV, it was useful to provide a more detailed treatment of some particular geometries, in view of the not so transparent algebra involved in Eq. (\ref{gen}). Nevertheless, the power of the general dispersion relation is that it gives the perturbation of Trivelpiece-Gould modes by neutrino effects for arbitrary angular orientation of wave vector, neutrino beam and ambient magnetic field. 

To enhance the neutrino contribution in Eq. (\ref{gen}) one has $\omega \approx {\bf k}\cdot{\bf u}_{\nu 0} \ll c k$. At the same time, Landau damping is relevant for resonant particles with $\omega - l \omega_c - k_z v_z \approx 0$, where $k_z = k\cos\theta$. To avoid this in the case of the fundamental mode ($l = 0$) one then needs $k << \omega/(v_T \cos\theta)$ or just $k << \omega/v_T$, for simplicity and similarly to the previous choices. In this context, as before we set the wave-number $k = \omega/v_T$, similarly to Eq. (\ref{lel}), with the understanding that the obtained growth rate estimate is the upper limit of it.

It can be verified that neutrino beam velocities compatible with $|{\bf u}_{\nu 0}| \approx c \gg v_T = \omega/k = {\bf k}\cdot{\bf u}_{\nu 0}/k$ are given by 
\begin{equation}
{\bf u}_{\nu 0} = (v_T \sin\theta + c\cos\alpha\cos\theta, c \sin\alpha, v_T \cos\theta - c \cos\alpha\sin\theta) \,,
\end{equation}
where $\alpha$ is an arbitrary angle. Setting $\omega = \omega_{\pm} + \delta$, where $|\delta| \ll \omega_{\pm}$ and where 
\begin{equation}
\omega_{\pm}^2 = \frac{1}{2}(\omega_H^2 \pm \Omega^2) \,, \quad \Omega^2 = \left((\omega_p^2 - \omega_c^2)^2 + 4 \omega_p^2 \omega_c^2 \sin^{2}\theta\right)^{1/2} 
\end{equation}
gives the unperturbed frequencies and working as before, the unstable root with ${\rm Im}(\delta) > 0$ is found with 
\begin{eqnarray}
{\rm Im}(\delta) &=& \frac{\sqrt{3}\,\Delta^{1/3}}{2^{4/3} |\omega_{\pm}^2 - \omega_c^2|^{1/3} \omega_{\pm}^{1/3} \Omega^{2/3}} \times \nonumber \\ 
&\times& \Bigl\{\Bigl[ \omega_{\pm}^2 \Bigl((\omega_{\pm}^2 - \omega_{c}^2)c^2/v_T^2 - \omega_{p}^2\Bigr) + \omega_p^2 \omega_c^2 \cos\theta (\cos\theta - (c/v_T)\cos\alpha\sin\theta)\Bigr]\omega_{\pm}^2/\omega_p^2 + \nonumber \\
&+& \omega_c^2 \omega_p^2 \omega_{\pm}^4 (c^2/v_T^2) \sin^{2}\theta\sin^{2}\alpha\Bigr\}^{1/3} \,. \label{comp}
\end{eqnarray}

It turns out that the choice of $\alpha$ is not numerically relevant for realistic physical estimates. Setting $\alpha = 0^{\circ}$, using the non-relativistic assumption $v_T^2/c^2 \ll 1$ and replacing the zero order dispersion relation $\omega_p^2 \omega_c^2 \cos^{2}\theta = \omega_H^2 \omega_{\pm}^2 - \omega_{\pm}^4$ whenever convenient allows to simplify Eq. (\ref{comp}) as 
\begin{equation}
{\rm Im}\Bigl(\frac{\delta}{\omega_\pm}\Bigr) = \sqrt{3} \Delta^{1/3} \left(\frac{\omega_{\pm}^2 |\omega_{\pm}^2 - \omega_c^2|}{\omega_p^2 \Omega^2}\right)^{1/3} 
\left(\frac{c}{2 v_T}\right)^{4/3} \,. \label{main}
\end{equation}
Equation (\ref{main}) is our final general result. When ${\bf k} \perp \boldsymbol{\omega}_c$ and $\omega_{\pm}^2 = \omega_{+}^2 \approx \omega_H^2$, it reproduces Eq. (\ref{insta}), while for $\omega_{\pm}^2 = \omega_{-}^2 \approx 0$ one has $\delta \approx 0$, justifying the neglect of the zero frequency mode in Section IVa. On the other hand, when ${\bf k} \parallel \boldsymbol{\omega}_c$ and $\omega_{\pm}^2 \approx \omega_p^2$, it reproduces Eq. (\ref{lel}), while setting $\omega_{\pm}^2 \approx \omega_{c}^2$ gives $\delta \approx 0$, which is in accordance with Section IVb where $\omega^2 \approx \omega_c^2$ was observed to be associated with zero neutrino density fluctuations. Notice that all neutrino effects shows up with the multiplicative factor $\Delta^{1/3} \sim G_F^{2/3}$. 

For some numerical estimates and for comparison we set the same parameters of the previous Sections, namely $n_0 = 10^{34} \,{\rm m}^{-3},  B = 5 \times 10^7 \, {\rm T}, v_T = c/10, {\cal E}_{\nu 0} = 10 \,{\rm MeV}$ with a prescribed equilibrium neutrino number density $n_{\nu 0} = 10^{35} \,{\rm m}^{-3}$ but keeping $\theta$ free, allowing a detailed observation of the dependence of the growth rate on the angle. The results are shown in Figs. \ref{figure4} and \ref{figure5} below, applying respectively for $\omega_-$ and $\omega_+$. In particular, in Fig. \ref{figure4} for $\theta = \pi/2 \,\,{\rm rad}$ (perpendicular propagation) gives $\delta \approx 0$ corresponding to $\omega^2 = \omega_{-}^2 = 0$. Similarly, In particular, in Fig. \ref{figure5} for parallel propagation gives $\delta \approx 0$ corresponding to $\omega^2 = \omega_{+}^2 = \omega_{c}^2 > \omega_{p}^2 = \omega_-^2$ for the chosen parameters. Finally, it can be verified that using the more general expression (\ref{comp}) also allowing the angle $\alpha$ to vary does not appreciably change the qualitative and quantitative findings. 

\begin{figure}[!hbt]
\begin{center}
\includegraphics[width=8.0cm,height=6.0cm]{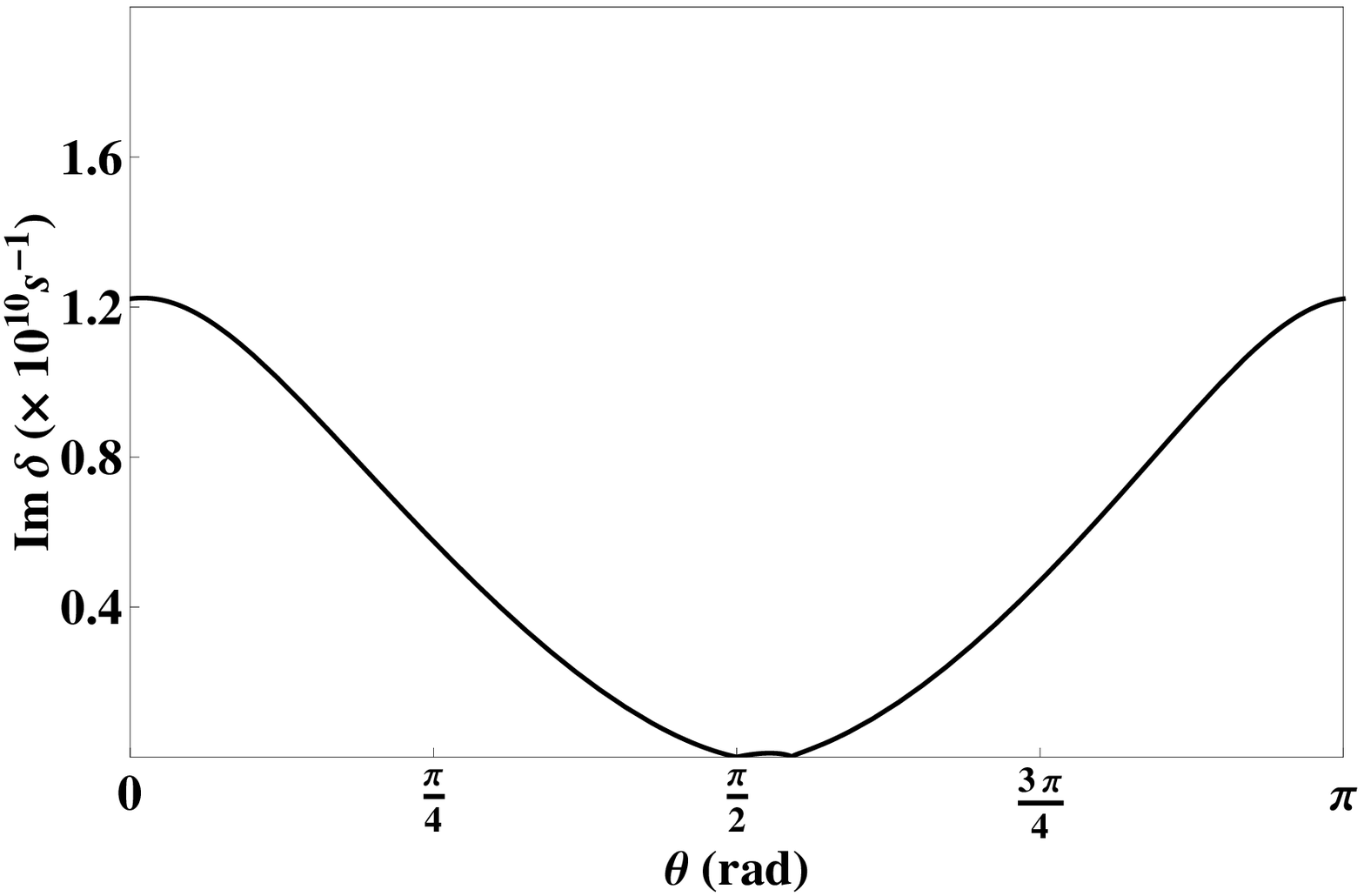}
\end{center}
\caption{Instability growth rate from Eq. (\ref{main})  as a function of $\theta$, using the mode $\omega_-$, for $n_0 = 10^{34} \, {\rm m}^{-3}, n_{\nu 0} = 10^{35} \, {\rm m}^{-3}, B = 5 \times 10^7 \, {\rm T}, v_T = c/10, {\cal E}_{\nu 0} = 10 \,{\rm MeV}$.}
\label{figure4}
\end{figure} 
\begin{figure}[!hbt]
\begin{center}
\includegraphics[width=8.0cm,height=6.0cm]{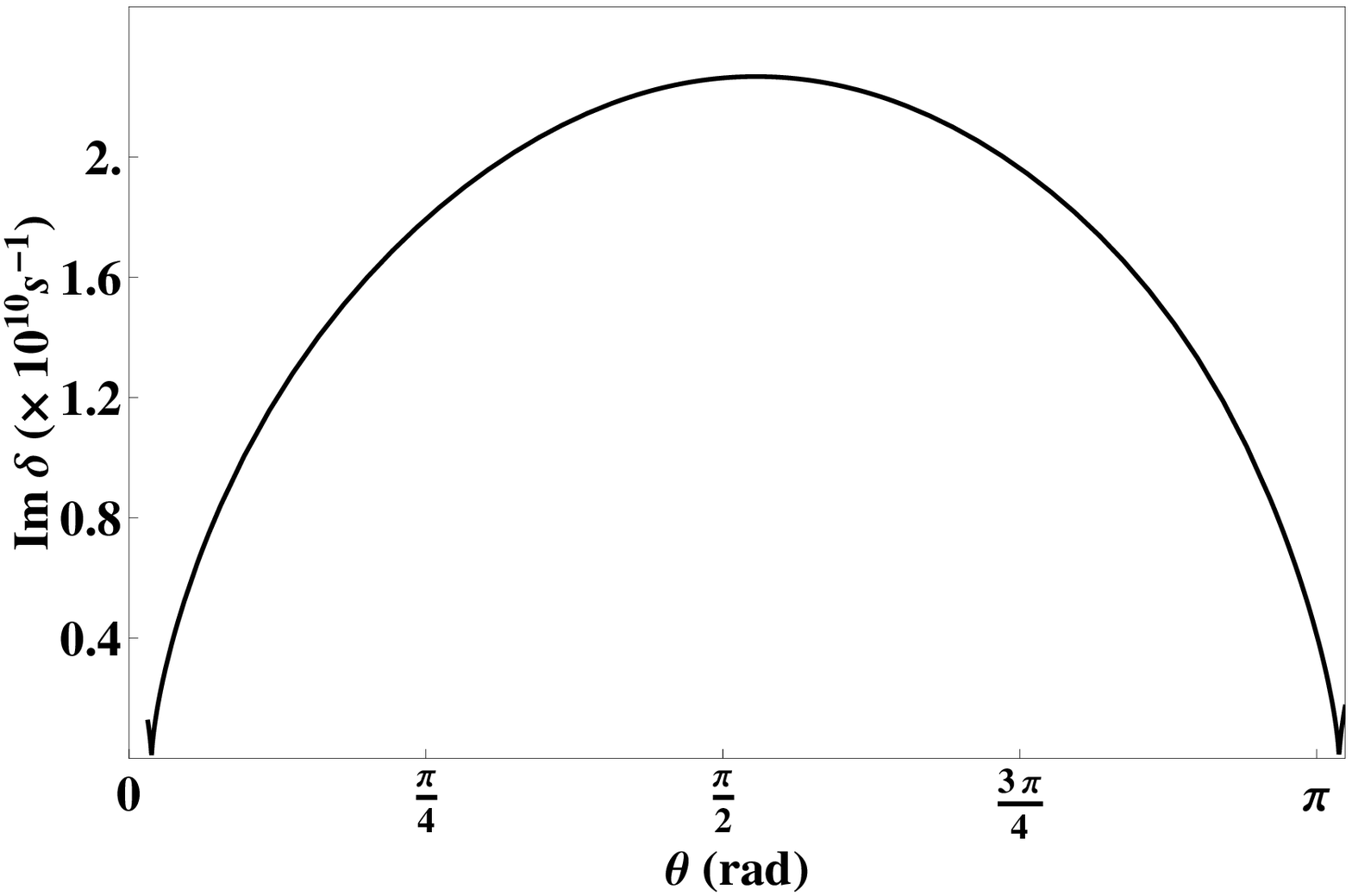}
\end{center}
\caption{Instability growth rate from Eq. (\ref{main})  as a function of $\theta$, using the mode $\omega_+$, for $n_0 = 10^{34} \, {\rm m}^{-3}, n_{\nu 0} = 10^{35} \, {\rm m}^{-3}, B = 5 \times 10^7 \, {\rm T}, v_T = c/10, {\cal E}_{\nu 0} = 10 \,{\rm MeV}$.}
\label{figure5}
\end{figure} 

\section{Conclusion}

In this work the destabilization of Trivelpiece-Gould modes due to interaction with a neutrino burst was established. The growth rate in dense magnetized plasma under intense neutrino beams was found to be significant, as in the case of conditions near the core of magnetized supernovae. It is found that the ambient magnetic field can enhance the instability, as in the case of perpendicular propagation where the essential result is the replacement of the plasma frequency by the upper hybrid frequency as the natural inverse time scale of the instability. 
The very general growth rate (\ref{main}) can be used to the analysis of neutrino-plasma interactions in a magnetized medium, in empirical tests of our understanding of the coupling between charged leptons and neutrinos. In particular, a complete treatment of the angular orientations of wave vector, neutrino beam and magnetic field is necessary for the plasma diagnostics and accuracy of the proposed model. Finally, the electron cyclotron Landau damping would be accessible by means of a kinetic treatment.

\acknowledgments
F.~H.~ and J.~T.~M.~ acknowledge the support by Con\-se\-lho Na\-cio\-nal de De\-sen\-vol\-vi\-men\-to Cien\-t\'{\i}\-fi\-co e Tec\-no\-l\'o\-gi\-co (CNPq) and EU-FP7 IRSES Programme (grant 612506 QUANTUM PLASMAS FP7-PEOPLE-2013-IRSES), and K.~A.~P.~ack\-now\-ledges the support by Coordena\c{c}\~ao de Aperfei\c{c}oamento de Pessoal de N\'{\i}vel Superior (CAPES). 

\appendix

\section{Full expressions of $\delta{\bf u}_\nu$ and $\delta n_\nu$}

Following the procedure outlined in Section III assuming $\omega^2 \neq \omega_c^2$ we get
\begin{eqnarray}
\delta{\bf u}_\nu &=& \frac{\sqrt{2} G_F \delta n_e c^2}{{\cal E}_{\nu 0}(\omega - {\bf k}\cdot{\bf u}_{\nu 0})(\omega^2 - \omega_{c}^2)} \times \Bigl[(\omega^2 - \omega_c^2 - \frac{\omega^2 \omega_p^2}{c^2 k^2}){\bf k} + \frac{\omega_p^2}{c^2 k^2} {\bf k}\cdot\boldsymbol{\omega}_c\,\boldsymbol{\omega}_c + \nonumber \\ 
&+&  \frac{\omega_p^2}{c^2 k^2} \frac{{\bf k}\cdot\boldsymbol{\omega}_c}{\omega}\,{\bf u}_{\nu 0}\times({\bf k}\times\boldsymbol{\omega}_c) - \frac{i \omega_p^2}{c^2 k^2} \omega \boldsymbol{\omega}_c \times{\bf k} - \frac{i \omega_p^2}{c^2 k^2}[{\bf u}_{\nu 0}\cdot(\boldsymbol{\omega}_c \times{\bf k})]{\bf k} + \nonumber \\
&+& \frac{i \omega_p^2}{c^2 k^2} ({\bf k}\cdot{\bf u}_{\nu 0}) (\boldsymbol{\omega}_c \times{\bf k}) - \frac{{\bf u}_{\nu 0}}{c^2} \,\, \Bigl((\omega^2 - \omega_c^2 - \frac{\omega^2 \omega_p^2}{c^2 k^2})({\bf k}\cdot{\bf u}_{\nu 0}) + \nonumber \\ &+& \frac{\omega_p^2}{c^2 k^2}({\bf k}\cdot\boldsymbol{\omega}_c)({\bf u}_{\nu 0}\cdot\boldsymbol{\omega}_c) - \frac{i \omega_p^2}{c^2 k^2} \omega {\bf u}_{\nu 0}\cdot(\boldsymbol{\omega}_c\times{\bf k})\Bigr)\Bigr] \,. \label{a1}
\end{eqnarray}
Then from the neutrino continuity equation we get
\begin{eqnarray}
\delta n_\nu &=& \frac{\sqrt{2} G_F n_{\nu 0} \delta n_e c^2}{{\cal E}_{\nu 0}(\omega - {\bf k}\cdot{\bf u}_{\nu 0})^2(\omega^2 - \omega_{c}^2)} \times \Bigl[(\omega^2 - \omega_c^2)k^2 - \frac{\omega^2 \omega_p^2}{c^2} + \frac{\omega_p^2}{c^2 k^2}({\bf k}\cdot\boldsymbol{\omega}_c)^2 + \nonumber \\
&+& \frac{\omega_p^2}{c^2 k^2}\frac{({\bf k}\cdot\boldsymbol{\omega}_c)}{\omega}\,{\bf k}\cdot[{\bf u}_{\nu 0}\times({\bf k}\times\boldsymbol{\omega}_c)] - \frac{i\omega_p^2}{c^2}[{\bf u}_{\nu 0}\cdot(\boldsymbol{\omega}_c\times{\bf k})] \label{a2} \\
- \frac{({\bf k}\cdot{\bf u}_{\nu 0})}{c^2}\Bigl((\omega^2 &-& \omega_c^2 - \frac{\omega^2 \omega_p^2}{c^2 k^2})({\bf k}\cdot{\bf u}_{\nu 0}) + \frac{\omega_p^2}{c^2 k^2}({\bf k}\cdot\boldsymbol{\omega}_c)({\bf u}_{\nu 0}\cdot\boldsymbol{\omega}_c) - \frac{i \omega_p^2}{c^2 k^2} \omega {\bf u}_{\nu 0}\cdot(\boldsymbol{\omega}_c \times{\bf k})
\Bigr) \Bigr] \,. \nonumber
\end{eqnarray}
Both expressions are needed to evaluate the neutrino contribution in the full dispersion relation shown in Eq. (\ref{ld}).

\end{document}